\documentclass[twocolumn]{book}
\usepackage[dvips]{graphicx,color}
\usepackage{makeidx,universe}

\makeauthorindex

\BookTitle{Proceedings of the XXIX PHYSICS IN COLLISION}

\CopyRight{\copyright 2009 by Universal Academy Press, Inc.}

\newcommand{\mygev}{\;\rm{Ge}\kern-0.06667em\rm{V\/}}
\newcommand{\mytev}{\;\rm{Te}\kern-0.06667em\rm{V\/}}
\newcommand{\mypt}{p_{\rm T}} 
\newcommand{\myet}{{E}_{\rm T}} 
\def\myetmiss{\mbox{\ensuremath{\, \slash\kern-.6emE_{T}}}}

\newcommand{\mypb}{\rm{pb}}

\newcommand{\FB}{$\rm\ fb^{-1}$}
\newcommand{\PB}{$\rm\ pb^{-1}$}
\newcommand{\ttbar}{$t\bar{t}$}

\begin{document} 

\pagenumbering{arabic}

\chapter{%
  {\LARGE \sf
    Single-Top Cross Section Measurements at ATLAS \\ Proceedings } \\
  {\normalsize \bf 
    Patrick Ryan$^1$ on behalf of the ATLAS collaboration } \\
  {\small \it \vspace{-.5\baselineskip}
    (1) Michigan State University, 
    Michigan State University, East Lansing, MI 48824, USA \\
  }
}




  \baselineskip=10pt 
  \parindent=10pt    

\section*{Abstract} 

The single-top production cross section is one third that of the top-pair
production cross section at the LHC.  During the first year of data taking, 
the determination of the major contributions to the total single-top cross 
section should be achievable. Comparisons between the measured cross sections 
and the theoretical predictions will provide a crucial test of the standard model.  
These measurements should also lead to a direct measurement of $|V_{tb}|$ with a 
precision at the level of a few percent. In addition, they will probe for new physics 
via the search for evidence of anomalous couplings to the top quark and measurements 
of additional bosonic contributions to single-top production. Methods developed to
optimize the selection of single-top events in the three production channels are
presented and the potential for the cross section measurements is established.

%
%

\section{Single-Top Physics}
\begin{figure}[b]
  \begin{center}
    \begin{tabular}{ccc}
      
      \begin{minipage}{2.5cm}
        \centering
        \includegraphics[height=2.5cm,width=2.5cm]{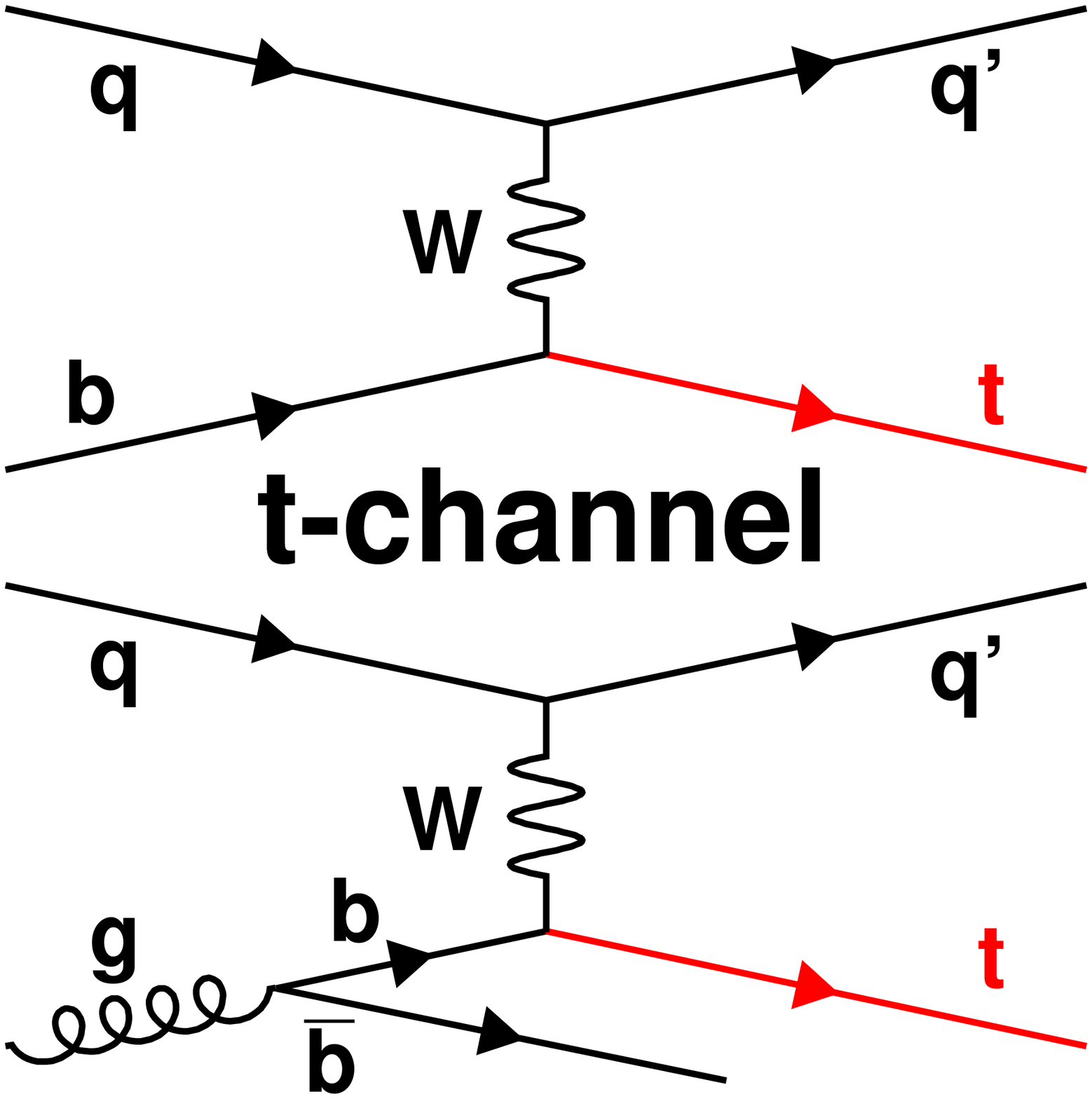}
      \end{minipage}
      
      \begin{minipage}{2.5cm}
        \centering
        \includegraphics[height=2.5cm,width=2.5cm]{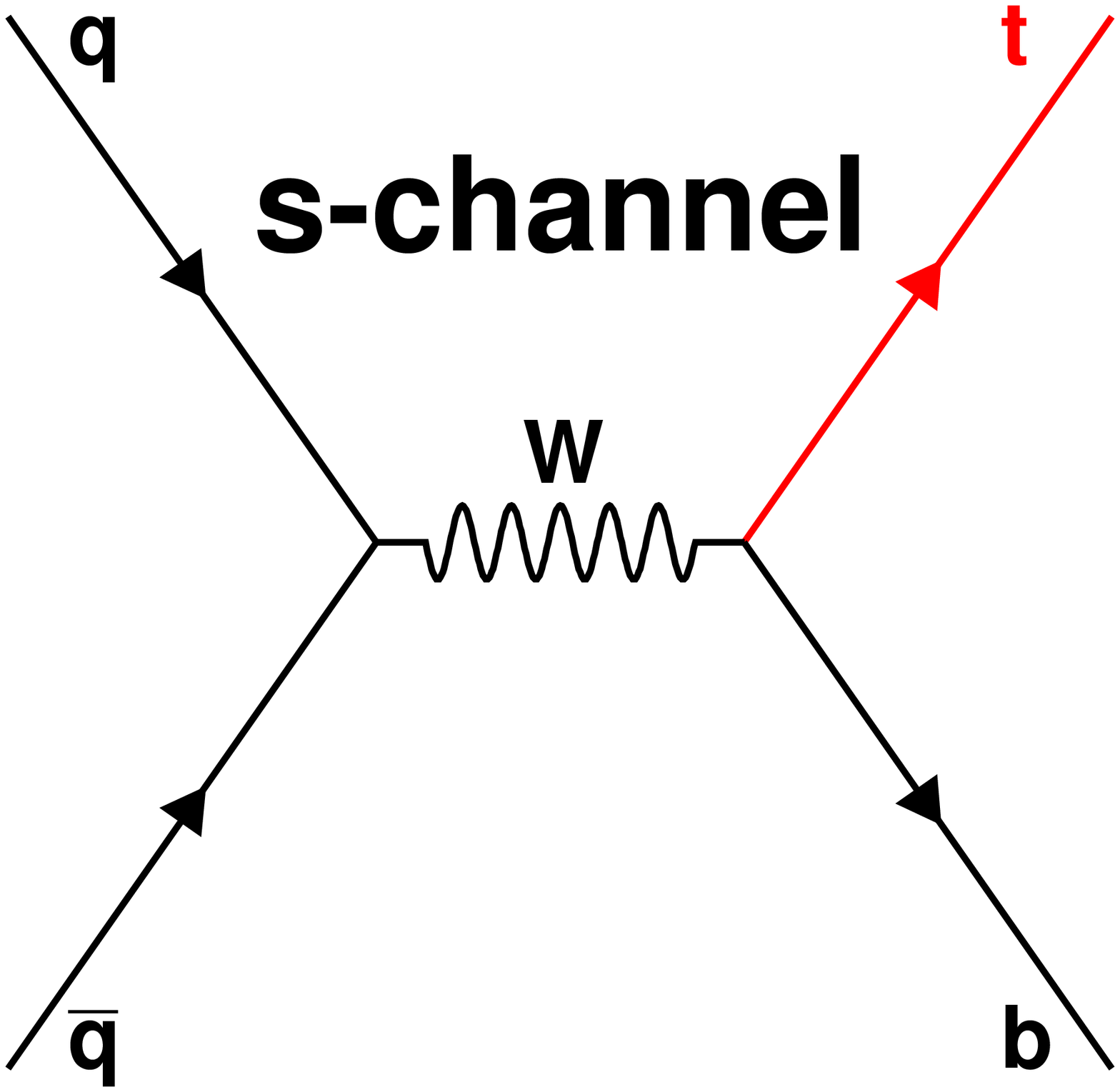}
      \end{minipage}

      \begin{minipage}{2.5cm}
        \centering
        \includegraphics[height=2.5cm,width=2.5cm]{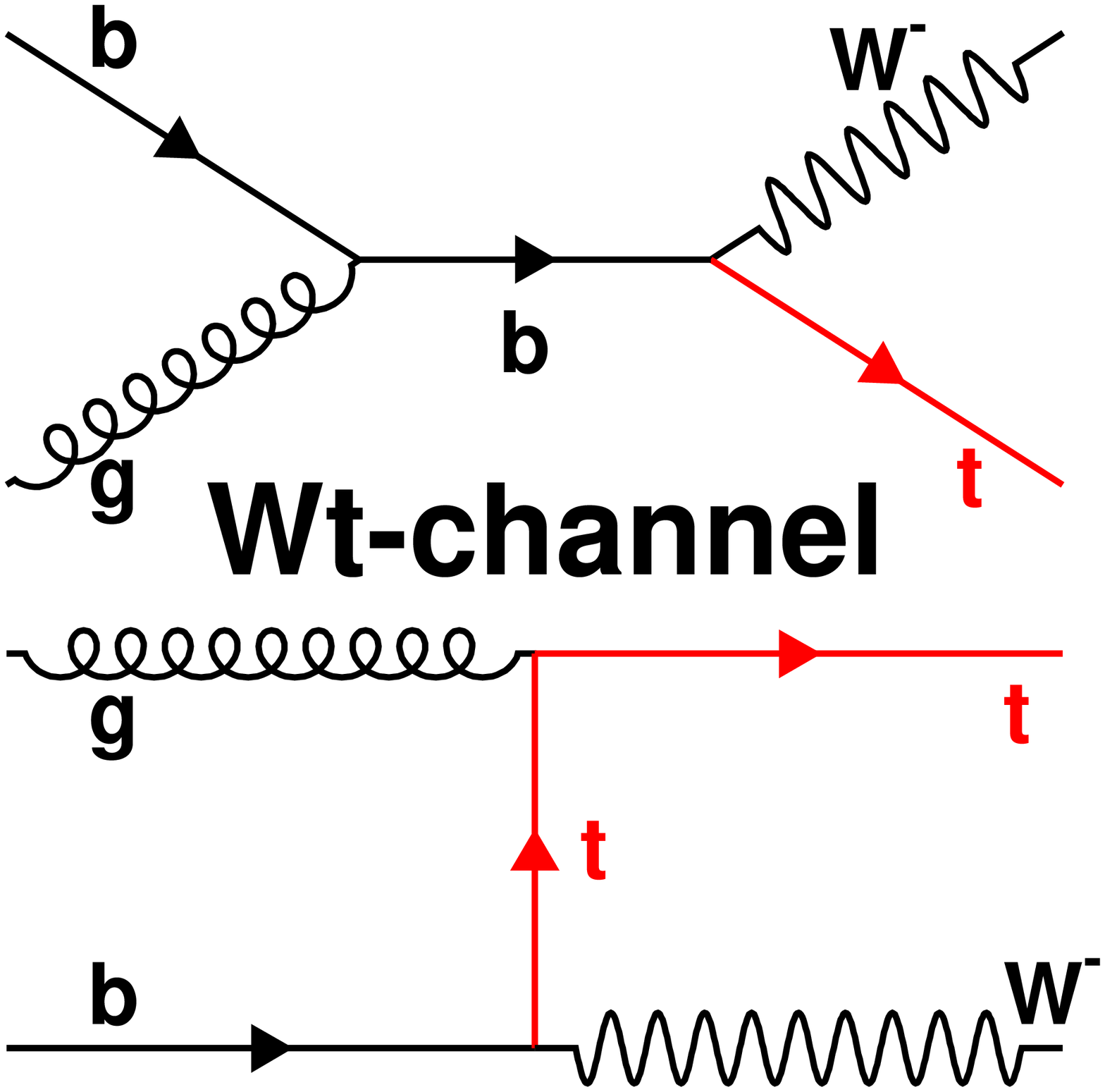}
      \end{minipage}
      
    \end{tabular}

    \caption[Single-top production diagrams.]
            {Single top production for t-channel (left), s-channel (middle), and $Wt$-channel (right).}
    \label{fig:singletop}
  \end{center}
\end{figure}

Single-top quark production was recently observed for the first time at Fermilab, where DZero measured
a cross section of $3.94 \pm 0.88\mypb$~\cite{D0} and CDF measured a cross section of $2.3^{+0.6}_{-0.5}\mypb$~\cite{CDF}.
Single-top quarks are produced via the electroweak interaction and at leading order there are three
different production mechanisms: s-channel, t-channel, and $Wt$-channel, each of which are depicted in
Figure~\ref{fig:singletop} 
At $14\mytev$, the theoretical cross sections for the single-top processes are $246 \pm 10.2 \mypb$~\cite{Sullivan,Campbell} for the t-channel,
$10.65 \pm 0.65 \mypb$~\cite{Sullivan,Campbell} for the s-channel, 
and $66.5 \pm 3.0 \mypb$~\cite{Campbell2} for the $Wt$-channel.
Only single-top events with an isolated and high-$\mypt$ muon or electron in the final state
are included in this study.  Single-top events with an all-hadronic final state are excluded.

The dominant background to single-top production is \ttbar\ production, which has a cross section of $833^{+52}_{-39} \mypb$~\cite{Bonciani}.
With a single high-$\mypt$ lepton, two $b$-jets, and $\myetmiss$, semi-leptonic \ttbar\ decay is most likely to mimic 
single-top events.
$W$ + jets production constitutes a significant background
since the cross-sections for these processes are approximately two orders of magnitude greater than the
single-top cross sections.  
The background from di-boson events is approximately three orders of magnitude smaller than the single-top
signal cross sections and therefore contributes minimally to the background.
QCD will be estimated by data driven methods and is not considered in these studies.  
The amount of QCD contamination depends on the specific selections used in the analyses.

%
%
\section{Single-Top Pre-selection}
The three single-top processes shared a common pre-selection.  Muons and electrons were required to satisfy
reconstruction requirements of $\myet > 10\mygev$ and $|\eta| < 2.5$ and an isolation requirement
of $\myet < 6\mygev$ in a cone of radius 0.2 around the particle axis.
Events were required to contain one muon or electron with $\mypt > 30\mygev$ and events with secondary leptons
were removed to eliminate contamination from di-lepton \ttbar\ events and to ensure the orthogonality 
of the muon and electron samples.
Jet candidates were reconstructed using a cone algorithm with $\Delta R = 0.4$ and were required to satisfy
$\mypt > 15\mygev$ to be considered a jet.  An event was required to have between 2 and 4 jets, with
at least two of the jets having $\mypt > 30\mygev$ and at least one of the jets being $b$-tagged.  
Single-top events have between one and three jets in the final state.
Events were required to have $\myetmiss > 25 \mygev$,
which allows for the detection of leptonic $W$ decays.  The trigger selection required muons to have $\mypt > 10\mygev$
and isolated electrons to have $\mypt > 25\mygev$ or non-isolated electrons with $\mypt > 60\mygev$.

%
%
\section{Cross Section Measurements}
The measurement of the single-top cross sections in the ATLAS detector~\cite{ATLAS-Detector}
will be obtained using the formula
$$\sigma = \frac{N_{Data} - N_{bkg}}{\epsilon_{S} \times \cal{L}},$$
where $N_{Data}$ is the total number of events in the data, $N_{bkg}$ is the
number of expected background events, $\epsilon_{S}$ is the selection efficiency
for single-top signal events, and $\cal{L}$ is the luminosity.

The sources of experimental uncertainty considered in the analysis were
Jet Energy Scale (JES), $b$-tagging likelihood, and luminosity.
The sources of theoretical uncertainty were background cross sections, 
Initial State Radiation (ISR) and Final State Radiation (FSR),
PDFs, and $b$ quark fragmentation.

Cut-based and multivariate analyses of the cross section measurements 
were performed for each of the three single-top channels in order to
understand the size of the statistical and systematic errors and the amount
of integrated luminosity needed to obtain evidence for and achieve discovery of the 
single-top quark~\cite{CSC-Note}.  For each single-top channel, the multivariate analysis improved sensitivity
compared to the cut-based analysis.

The cut-based analysis for the t-channel was performed by requiring, in addition to the
event pre-selection, a $b$-tagged jet with $\mypt > 50\mygev$ in order to remove
low-$\mypt$ $W$ + jets background and $|\eta| > 2.5$ for the hardest light
jet to remove \ttbar\ contamination.  
The presence of a large number of \ttbar\ events remaining after the t-channel specific event selection
necessitated the use of a Boosted Decision Tree (BDT), which maximized \ttbar\ separation and signal over background.

\begin{figure}
  \begin{center}
    \includegraphics[height=4.5cm,width=7.38cm]{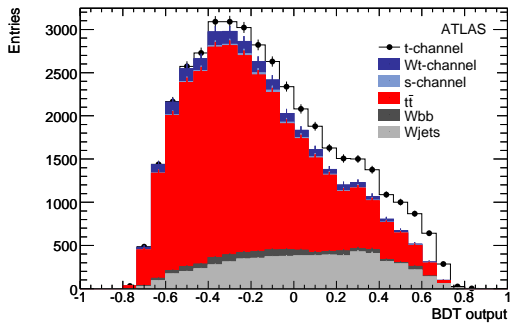}
    
    \caption[t-channel BDT output.]
            {The BDT output for the t-channel.}
    \label{fig:t-channel:BDT}
  \end{center}
\end{figure}

The set of variables used in the BDT was chosen in order to reduce the impact of JES systematics as much as possible
while retaining a large discrimination power.
The BDT output is shown in Figure~\ref{fig:t-channel:BDT}
A cut on the BDT output was set so that the total (statistical + systematic) uncertainties were minimized.
After such a selection, the resulting signal over background ratio was 1.3.  The dominant background
contributions were from \ttbar\ and the other single-top processes.

%
%
%
%
The statistical uncertainties for 1\FB of luminosity were 5.0\% for the cut-based analysis and 5.7\% for the BDT analysis.
The total uncertainties were 45\% for the cut-based analysis and 23\% for the BDT analysis.
The systematic uncertainties far outweighed the statistical uncertainties in both analyses.
The main systematic contributions were JES, ISR and FSR, and luminosity.  

The t-channel cross section is proportional to $|f_{L}V_{tb}|^2$, where $f_L$ is
the weak left-handed coupling and equal to 1 in the Standard Model.  The estimated
uncertainty on $|V_{tb}|$ was calculated to be
${\Delta\left|V_{tb}\right|}/{\left|V_{tb}\right|} = \pm 11.2\%_{stat+sys} \pm 3.9\%_{theory} = \pm 11.9\%.$

The s-channel cut-based analysis was performed by requiring exactly two jets to account for 
event topology, which rejected mostly \ttbar\ events.  The two jets were required to
be $b$-tagged in order to reject $W$ + jets and QCD background, both of which are characterized
by soft $b$-jets or no $b$-jets at all.  
The high background levels remaining after the cut-based analysis motivated the use of likelihood
functions (LFs), which set thresholds by minimizing the total uncertainty.
For 10\FB of luminosity, the statistical uncertainty was 20\% and the total uncertainty was 52\%, from which it can be concluded
that the cross section measurement is both statistically and systematically limited.
The dominant systematics were ISR and FSR modeling, knowledge of the background cross sections, and luminosity determination.  

%
%

\begin{figure}
  \begin{center}
    \includegraphics[height=5.3cm,width=5.49cm]{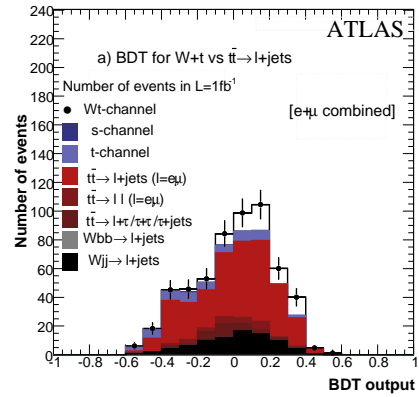}
    \caption[$Wt$-channel BDT output \ttbar\ $\rightarrow$ lepton + jets.]
            {The $Wt$-channel BDT output for \ttbar\ $\rightarrow$ lepton + jets.}
    \label{fig:Wt-channel:BDT}
  \end{center}
\end{figure}

The $Wt$-channel cut-based analysis required one $b$-tagged jet with $\mypt > 50\mygev$ to account
for event topology.  In addition, events having additional $b$-tagged jets
with $\mypt > 35\mygev$ were vetoed to reject \ttbar\ events.  
Compared to the b-jets upon which the $\mypt$ requirement was imposed, 
those used in the veto were selected with a looser b-tag weight
which was optimized according to the signal over \ttbar\ background ratio.  
BDTs were used to discriminate the $Wt$-channel signal against \ttbar\ events in the lepton + jets 
and di-lepton channels, $W$ + jets, and t-channel events.  Discrimination for lepton + jets is shown in Figure~\ref{fig:Wt-channel:BDT}
BDT thresholds were set by minimizing the total uncertainty.  
For 1\FB\ of luminosity, the statistical uncertainty was 21\% and the overall uncertainty was 52\%.
Systematics dominate the measurement and the most dominant systematics were ISR and FSR, background cross sections, and luminosity.
%
%
\section{Conclusions}
%
Rediscovery of the single-top processes at the LHC will require several hundred \PB\ of well-understood data. 
Cross-section measurements with early-data will be dominated by the systematic uncertainties associated with b-tagging and JES determination.

Initial LHC running is forecasted to be at $7\mytev$, which would lead to approximately $7,500$ single-top events
with 100\PB\ of integrated luminosity.  This amount is similar to the number of single-top events in the Tevatron single-top discovery analysis
and should allow limits to be set on all single-top processes and the ruling out of Standard Model t-channel at 95\%.
Data driven techniques will be used as much as possible because of the large theoretical uncertainties. It is planned to use
multivariate techniques to estimate background using orthogonal \ttbar\ and $W$+ jets samples.

Single-top t-channel and $Wt$-channel discoveries at $14\mytev$ are expected to require approximately 1 \FB\ of luminosity and precision measurements
are expected to require a few \FB.  Single-top s-channel discovery is expected to require approximately 10 \FB\ of luminosity and statistics, 
in addition to systematics, are foreseen to play a limiting role.


\begin{thebibliography}{99}   









\bibitem{D0} V.~M.~Abazov {\it et al.}  [D0 Collaboration], ``Observation of Single Top-Quark Production,'' Phys.\ Rev.\ Lett.\  {\bf 103} (2009) 092001.

\bibitem{CDF} T.~Aaltonen {\it et al.}  [CDF Collaboration], ``First Observation of Electroweak Single Top Quark Production,'' Phys.\ Rev.\ Lett.\  {\bf 103} (2009) 092002.


\bibitem{Sullivan}
Z.~Sullivan,
``Understanding single-top-quark production and jets at hadron colliders,'' Phys.\ Rev.\  D {\bf 70} (2004) 114012


\bibitem{Campbell}
J.~M.~Campbell, R.~K.~Ellis and F.~Tramontano, ``Single top production and decay at next-to-leading order,'' Phys.\ Rev.\  D {\bf 70} (2004) 094012


\bibitem{Campbell2}
J.~Campbell and F.~Tramontano, ``Next-to-leading order corrections to Wt production and decay,'' Nucl.~Phys~(2005) B726

\bibitem{Bonciani}
R.Briconi et al., ``NLL resummation of the heavy-quark hadroproduction cross-section,'' Nucl.~Phys~(1998) B529.

\bibitem{ATLAS-Detector}
G.~Aad et al., ``The ATLAS Experiment at the CERN Large Hadron Collider,'' JINST 3 (2008) S08003.

\bibitem{CSC-Note}
ATLAS Collaboration, ``Expected Performance of the ATLAS Experiment,'' Detector, Trigger and Physics CERN-OPEN-2008-020, Geneva, 2008. 


\end{thebibliography}
\end{document}